\documentclass[a4paper,11pt]{article}
\usepackage[margin=2.5cm]{geometry}
\usepackage[utf8]{inputenc}
\usepackage[T1]{fontenc}
\usepackage{hyperref}
\usepackage{amsmath,amsfonts,amssymb,array,graphicx,color}
\usepackage{authblk}

\graphicspath{{figpdf/}}

\makeatletter

\@addtoreset{equation}{section}
\makeatother
\newcommand{\appref}[1]{Appendix~\ref{sec:#1}}
\newcommand{\secref}[1]{Sec.~\ref{sec:#1}}
\newcommand{\figref}[1]{Fig.~\ref{fig:#1}}

\newcommand{\normal}[1]{\, :\! #1 \!:\, }
\newcommand{\aver}[1]{\left\langle {#1} \right\rangle}
\newcommand{\smallaver}[1]{\langle {#1} \rangle}

\newcommand{\ket}[1]{| {#1} \rangle}
\newcommand{\vect}[1]{\boldsymbol{#1}}
\newcommand{\kac}[2]{\left(\begin{array}{cc} #1 \\ #2 \end{array} \right)}

\newcommand{\id}{\mathbf{1}}

\newcommand{\nn}{\nonumber}
\newcommand{\zb}{\bar{z}}

\newcommand{\om}{\vect\omega}
\newcommand{\vQ}{\vect{Q}}
\newcommand{\ve}{\vect{e}}
\newcommand{\vh}{\vect{h}}
\newcommand{\valpha}{\vect{\alpha}}

\newcommand{\vlambda}{\vect{\lambda}}
\newcommand{\vmu}{\vect{\mu}}
\newcommand{\vnu}{\vect{\nu}}
\newcommand{\vrho}{\vect{\rho}}
\newcommand{\Zbb}{\mathbb{Z}}
\newcommand{\eps}{\varepsilon}
\newcommand{\wh}{\widehat}
\newcommand{\wt}{\widetilde}
\newcommand{\sla}{\mathfrak{sl}}

\title{Fusion of spin fields in $W_3$ conformal field theories}
\author{Yacine Ikhlef$^1$ and Hirohiko Shimada$^{2,3}$}
\affil{$^1$ Sorbonne Universit\'e, CNRS, Laboratoire de Physique Th\'eorique et Hautes \'Energies, LPTHE, F-75005 Paris, France}
\affil{$^2$ Mathematical and Theoretical Physics Unit, Okinawa Institute of Science and Technology
Graduate University, Onna, Okinawa, 904-0495, Japan}
\affil{$^3$ Department of Integrated Science and Technology, National Institute of Technology, Tsuyama College, 624-1 Numa, Tsuyama, Okayama 708-8509, Japan}

\date{\today}

\begin{document}

\maketitle

\begin{abstract}
  In generic conformal field theories with $W_3$ symmetry, we identify a primary field $\sigma$ with rational Kac indices, which produces the full $\mathbb{Z}_3$ charged and neutral sectors by the fusion processes $\sigma \times \sigma$ and $\sigma \times \sigma^*$, respectively. In this sense, this field generalises the $\mathbb{Z}_3$ fundamental spin field of the three-state Potts model. Among the degenerate fields produced by these fusions, we single out a ``parafermion'' field $\psi$ and an ``energy'' field $\varepsilon$. In analogy with the Virasoro case, the exact curves for conformal dimensions $(h_\sigma,h_\psi)$ and $(h_\sigma,h_\varepsilon)$ are expected to give close estimates for the unitarity bounds in the conformal bootstrap analysis.
\end{abstract}

\section{Introduction}
\label{sec:intro}

In the context of Conformal Field Theories (CFTs) describing the scaling limit of critical lattice models, the conformal bootstrap approach \cite{BPZ84,DotsenkoFateev84} has been a powerful tool to study the Operator Product Expansion (OPE) algebra, and to compute its structure constants (OPE coefficients). It is based on a few assumptions on correlation functions of primary operators: their decomposition into conformal blocks, their monodromy as one operator winds around another, and the discrete symmetries obeyed by fusion rules. More recently, this approach has been used successfully to design numerical algorithms for the study of critical models, including the case of dimension $d>2$ \cite{Ising3d,SAW3d}, or 2d correlation functions whose internal spectrum is not known \textit{a priori} \cite{RibaultSanta15}.
\bigskip

A particular application of the bootstrap approach consists in considering a ``fundamental'' spin field $\sigma$, which produces the full set of primary fields when iteratively fused with itself, and demanding that the corresponding OPE coefficients are real. For instance, in the case of the Ising model in any dimension $d$, the spin field $\sigma$ should be odd under $\Zbb_2$ symmetry, and should obey a fusion of the form :
\begin{equation*}
  \sigma \times \sigma \to \id + \eps + \dots
\end{equation*}
where $\eps$ is the most relevant non-trivial primary operator in the even sector.
The positivity of the squared OPE coefficient $C^2(\sigma,\sigma,\eps)$ within the bootstrap of the four-point function $\aver{\sigma\sigma\sigma\sigma}$ then leads to the determination of forbidden regions in the diagram of conformal dimensions $(h_\sigma,h_{\eps})$ \cite{Ising3d}. A similar approach can be applied to the O($n$) vector model \cite{bootstrap-On}. 
\bigskip

In the case of 2d CFTs governed by the Virasoro algebra, the spin field can be identified generically as a primary field with half-integer Kac indices $\sigma \equiv \Phi_{1/2,0}$ , and the fusion $\sigma \times \sigma$ produces the infinite series of ``energy-like'' operators \cite{DotsenkoFateev84}:
\begin{equation*}
  \sigma \times \sigma \to \id + \eps^{(1)} + \eps^{(2)} + \eps^{(3)} + \dots
\end{equation*}
The following features have been observed recently \cite{Shimada18} for this set of operators:
\begin{itemize}
\item Up to proper identification of the most relevant energy operator $\eps=\eps^{(1)}$ in the O($n$) loop model and the Fortuin-Kasteleyn cluster model (see \secref{Virasoro}), the lines $(h_\sigma,h_{\eps})$ for these two models give very good approximations to the unitarity bounds found by the numerical bootstrap.
\item The sequence of zeroes and poles for the squared OPE coefficients $C^2(\sigma,\sigma,\eps^{(k)})$ follow a structure which can be encoded in terms of Farey paths on the Poincar\'e disk, and the congruence subgroup $\Gamma(2)$ of SL($2,\Zbb$).
\end{itemize}
In this paper, our aim is to set up the bases for a similar study of OPE coefficients, in the case of an extended conformal symmetry governed by the $W_3$ algebra \cite{Zamo85,FateevZamo87,FateevLukyanov88}. In particular, we develop an argument for the proper identification of spin fields and energy-like operators, by reasoning especially on the $\Zbb_3$ symmetry arising in the representation theory of the $W_3$ algebra. A third class of operators, the $\Zbb_3$ parafermions $\psi$ and $\psi^*$, shall appear naturally in our study.
\bigskip

The structure of the paper is as follows. In \secref{W3-CFT} we review some useful background on $W_3$ CFTs,
% as found in~\cite{Zamo85,FateevZamo87,FateevLukyanov88},
and we discuss the conformal blocks of four-point correlation functions involving a completely degenerate field and its conjugate. In \secref{spin}, %%%%
we start with reviewing some important facts about the fundamental spin field in the Virasoro CFTs.
Then we state the criteria for the identification of spin fields in a $W_3$ CFT, we derive a set of fields $\sigma,\sigma',\sigma''$ meeting these criteria in a stronger or weaker sense, and we discuss of the phase diagrams $(h_\sigma,h_\psi)$ and $(h_\sigma,h_\eps)$, where $\psi$ and $\epsilon$ are primary fields appearing in the fusions $\sigma \times \sigma$ and $\sigma \times \sigma^*$, respectively. In \secref{concl}, we conclude with some perspectives. In \appref{sl3}, we gather the notations and basic notions of representation theory of $\sla_3$ which are relevant to the discussion.

\section{Conformal Field Theories with $W_3$ symmetry}
\label{sec:W3-CFT}

\subsection{The $W_3$ conformal algebra}
\label{sec:W3-alg}

The $W_3$ algebra is an extended conformal algebra based on the stress-energy tensor $T(z)$ and an additional current $W(z)$ of dimension three \cite{Zamo85,FateevZamo87,FateevLukyanov88}. The mode decomposition reads
\begin{equation}
  T(z) = \sum_{n=-\infty}^{+\infty} L_n \, z^{-n-2} \,,
  \qquad W(z) = \sum_{n=-\infty}^{+\infty} W_n \, z^{-n-3} \,,
\end{equation}
and the commutation relations between the modes are:
\begin{equation} \label{eq:comm}
  \begin{aligned}
    {[L_n,L_m]} &= (m-n)L_{n+m} + \frac{c}{12}(n^3-n) \delta_{n+m,0} \,, \\
    {[L_n,W_m]} &= (2n-m) W_{n+m} \,, \\
    {[W_n,W_m]} &= \frac{c}{3\times 5!}(n^2-4)(n^3-n) \delta_{n+m,0}
    + \beta^2(n-m) \Lambda_{n+m}  \\
    & \qquad + (n-m)\left[ \frac{1}{15}(n+m+2)(n+m+3)
      - \frac{1}{6}(n+2)(m+2) \right] L_{n+m} \,,
  \end{aligned}
\end{equation}
where
\begin{equation}
  \begin{aligned}
    \beta &= \sqrt{\frac{16}{22+5c}} \,, \\
    \Lambda_n &= \sum_{k=-\infty}^{+\infty} \normal{L_kL_{n-k}} + \frac{x_n}{5} L_n \,,
    \qquad \normal{L_nL_m} = \begin{cases} L_nL_m & \text{if $n \leq m$} \\ L_mL_n & \text{if $n > m$} \end{cases} \\
    x_{2\ell} &=(1+\ell)(1-\ell) \,, \qquad x_{2\ell+1} = (2+\ell)(1-\ell) \,.
  \end{aligned}
\end{equation}
A primary field $\Phi_{h,w}$ is a highest-weight state for the algebra:
\begin{equation}
  L_{n>0} \Phi_{h,w} = W_{n>0} \Phi_{h,w} = 0 \,,
  \qquad L_0 \Phi_{h,w} = h \, \Phi_{h,w} \,,
  \qquad W_0 \Phi_{h,w} = w \, \Phi_{h,w} \,.
\end{equation}

\subsection{Coulomb-Gas parameterisation for $W_3$ theories}

The Coulomb-Gas (CG) approach provides a convenient parameterisation of the primary fields. Since many results are expressed in terms of the $\sla_3$ Lie algebra, we refer to the \appref{sl3} for the conventions used throughout this paper. The key point of the CG approach is to interpret the currents $T(z)$ and $W(z)$ as deriving from the action~\cite{FateevZamo87} :
\begin{equation} \label{eq:A}
  \mathcal{A}[\vect\phi] = \int \frac{d^2x}{8\pi} \sqrt{|g|} \left[
    \partial_\mu\vect\phi \cdot \partial^\mu\vect\phi
    + 2i R(\vec x) \vect Q \cdot \vect\phi
    + \normal{e^{i\ve_1 \cdot \vect\phi/b}} + \normal{e^{i\ve_2 \cdot \vect\phi/b}}
    \right] \,,
\end{equation}
where $\vect\phi$ is a two-component scalar field, $R(x)$ is the scalar curvature, $\ve_1$ and $\ve_2$ are the simple roots, and the background charge $\vQ$ is in the direction of the Weyl vector $\vrho=\ve_1+\ve_2$ :
\begin{equation} \label{eq:Q}
  \vQ = (b^{-1}-b) \vrho \,.
\end{equation}
The central charge and the parameter $\beta$ in \eqref{eq:comm} are then given by
\begin{equation} \label{eq:cc}
  c = 2 - 12\vQ^2 \,,
  \qquad \beta = \frac{2}{\sqrt{8-15\vQ^2}} \,.
\end{equation}
The rational model $\mathcal{M}_{p,p'}$ is parameterised by two coprime integers $p,p'$:
\begin{equation}
  b = \sqrt{\frac{p}{p'}} \,,
  \qquad c = 2-\frac{24(p-p')^2}{pp'} \,.
\end{equation}
Any primary field $\Phi_{h,w}$ can be represented as a vertex operator $\Phi_{h,w} \equiv V_{\valpha} = \exp(i\valpha \cdot \vect\phi)$, with eigenvalues for $L_0$ and $W_0$:
\begin{equation}
  h_{\valpha}= \frac{1}{2} \valpha \cdot( \valpha-2\vQ) \,,
  % =\frac{1}{2} \left[(\valpha-\vQ)^2 - \vQ^2 \right] \,,
  \qquad w_{\valpha}= \beta\sqrt{3} \, \prod_{j=1}^3 \left[(\valpha-\vQ)\cdot \vh_j \right] \,.
\end{equation}
The Weyl group $W$ and the conjugation act as follows on vertex charges:
\begin{align}
  \forall x \in W \,,
  \qquad & x \star \valpha = \vQ + x(\valpha-\vQ) \,, \label{eq:x.alpha} \\
  &(\alpha_1 \om_1 + \alpha_2 \om_2)^* = \alpha_2 \om_1 + \alpha_1 \om_2 \,.\label{eq:alpha.star}
\end{align}
The eigenvalues of $L_0$ and $W_0$ are invariant under the action of the Weyl group, whereas $w_{\valpha}$ changes sign under conjugation:
\begin{align}
  \forall x \in W \,, \qquad
  &h_{x \star \valpha} = h_{\valpha}
  \quad \text{and} \quad w_{x \star \valpha} = w_{\valpha} \,,\\
  &h_{\valpha^*} = h_{\valpha} 
  \quad \text{and} \quad w_{\valpha^*} = -w_{\valpha} \,.
\end{align}
Hence, for any $x \in W$, we identify $V_{x \star \valpha} \equiv V_{\valpha}$. We shall write $x \star \valpha \equiv \valpha$ for short. Moreover, since $R_{\vh_2} \star (2\vQ-\valpha)=\valpha^*$, where $R_{\vh_2}$ is the reflection about $\vh_2$, and $R_{\vh_2} \in W$, one can identify $\valpha^* \equiv 2\vQ-\valpha$.

\subsection{Semi- and completely degenerate primary fields}

A primary field is completely degenerate if it has a two-dimensional space of primary descendants~\cite{FateevZamo87}. The corresponding vertex charges are of the form:
\begin{equation} \label{eq:kac}
  \valpha\kac{n_1 & m_1}{n_2 & m_2} = \left[(1-n_1)b^{-1} - (1-m_1)b \right] \om_1 + \left[(1-n_2)b^{-1} - (1-m_2)b \right] \om_2 \,,
\end{equation}
with $n_1,n_2,m_1,m_2$ positive integers, called the Kac indices. The corresponding dimension is
\begin{equation} \label{eq:h-kac}
  h\kac{n_1 & m_1}{n_2 & m_2} =
  \frac{\left[(n_1+n_2) b^{-1} -(m_1+m_2)b \right]^2}{4}
  + \frac{\left[(n_1-n_2)b^{-1}-(m_1-m_2)b \right]^2}{12} - (b^{-1}-b)^2 \,,
\end{equation}
and the eigenvalue for $W_0$ is
\begin{align}
  w\kac{n_1 & m_1}{n_2 & m_2}
  &= \frac{2\beta \sqrt{3}}{27} \left[(n_1-n_2)b^{-1}-(m_1-m_2)b \right]
  \left[(n_1+2n_2)b^{-1}-(m_1+2m_2)b \right] \nn \\
  & \qquad \times \left[(2n_1+n_2)b^{-1}-(2m_1+m_2)b \right] \,.
  \label{eq:w-kac}
\end{align}
We denote the associated primary field as
\begin{equation} \label{eq:Phi}
  \Phi\kac{n_1 & m_1}{n_2 & m_2} = \Phi_{\vlambda,\vmu} \,,
  \qquad \text{with} \quad \lambda_i=n_i-1 \,, \quad \mu_i=m_i-1 \,.
\end{equation}
\bigskip

A primary field $\Phi$ is semi-degenerate if it has a \textit{one}-dimensional space of primary descendants. For instance,  $\Phi$ is semi-degenerate at level one iff:
\begin{equation}
  L_1 \, \chi = W_1 \, \chi = 0 \,,
  \qquad \text{with} \quad 
  \chi \propto (L_{-1} + \# W_{-1}) \Phi \,.
\end{equation}
This corresponds to a vertex operator $\Phi=V_{\valpha}$, with a charge of the form $\valpha=\kappa \om_1$ with $\kappa \in \mathbb{R}$, or any of the $x \star (\kappa \om_1)$ with $x \in W$.
In terms of Kac indices, $\valpha$ is of the form:
\begin{equation}
  \valpha \equiv \valpha\kac{n_1 & m_1}{1 & 1} \,, \qquad \text{with} \quad n_1,m_1 \in \mathbb{R} \,.
\end{equation}

\subsection{Fusion rules in generic $W_3$ models}
\label{sec:fusion}

We consider a model with generic central charge, \textit{i.e.} where $b^2$ in \eqref{eq:Q} is not rational. We indicate the fusion rules of the primary operator algebra by writing
\begin{equation}
  \Phi_i \times \Phi_j \to \sum_k \mathcal{N}_{ij}^k \,.\, \Phi_k \,,
\end{equation}
where $\mathcal{N}_{ij}^k$ is a positive integer giving the multiplicity of the term $\Phi_k$ in the OPE of $\Phi_i$ with $\Phi_j$.
We call ``generic'' any vertex charge $\valpha$ which satisfies:
\begin{equation}
  \valpha \notin \bigcup\limits_{j=1,2,3} \mathbb{R} \vh_j
  + \left(b^{-1} \mathcal{R}^* + b \mathcal{R}^* \right) \,.
\end{equation}
Note that this excludes semi- and completely degenerate fields.
One has the fusion rule (see \cite{FateevLitvinov05,FateevLitvinov07,FateevLitvinov09}):
\begin{equation} \label{eq:Phi.V}
  \Phi_{\vlambda,\vmu} \times V_{\valpha} \to
  \sum_{\vlambda' \in [\vlambda],\ \vmu' \in [\vmu]}
  m_{\vlambda}(\vlambda')m_{\vmu}(\vmu')
  \,.\, V_{\valpha-b^{-1}\vlambda'+b\vmu'} \,,
\end{equation}
where $m_{\vlambda}(\vlambda'),m_{\vmu}(\vmu')$ are the weight multiplicities (see \appref{sl3}). The fusion rule of two completely degenerate fields has the form \cite{FateevZamo87}
\begin{equation} \label{eq:Phi.Phi}
  \Phi_{\vlambda,\vmu} \times \Phi_{\vlambda',\vmu'} \to
  \sum_{\vlambda'', \vmu''} N_{\vlambda\vlambda'}^{\vlambda''}
  \, N_{\vmu\vmu'}^{\vmu''}
  \,.\, \Phi_{\vlambda'',\vmu''} \,,
\end{equation}
where
%\begin{align*}
%  & \lambda_j=n_j-1 \,, \qquad \lambda'_j=n'_j-1 \,, \qquad \lambda''_j=n''_j-1 \,, \\
%  & \mu_j=m_j-1 \,, \qquad \mu'_j=m'_j-1 \,, \qquad \mu''_j=m''_j-1 \,,
%\end{align*}
%and
$N_{\vlambda\vlambda'}^{\vlambda''}$ and $N_{\vmu\vmu'}^{\vmu''}$ are the fusion coefficients of $\sla_3$ representations.

\subsection{Four-point conformal blocks}

Let $\Phi = \Phi_{\vlambda,\vmu}$ be a completely degenerate primary field, parameterised by two $\sla_3$ heighest weights $\vlambda$ and $\vmu$ : see \eqref{eq:Phi}. Let $V_{\valpha}$ be a generic vertex operator (see \secref{fusion}). We consider the correlation function:
\begin{equation} \label{eq:G(z,zb)}
  G(z,\zb) = \aver{\Phi_{\vlambda,\vmu}^*(\infty) V_{\valpha}^*(1) V^{\phantom{*}}_{\valpha}(z,\zb) \Phi^{\phantom{*}}_{\vlambda,\vmu}(0)} \,.
\end{equation}
We discuss here the (unnormalised) $W_3$ conformal blocks associated to $G(z,\zb)$, in the channels $z \to 0$ and $z \to 1$, respectively:
\begin{align}
  \mathcal{F}_{i,a,b}(z) &= \sum_m z^{-h_\Phi-h_{\valpha}+h_i^{(m)}}
  {\smallaver{\Phi_{\vlambda,\vmu}^*| V_{\valpha}^*(1)|\Phi_i^{(m)}}_a
    \, \smallaver{\Phi_i^{(m)}|V_{\valpha}(1) |\Phi_{\vlambda,\vmu}}_b} \,,
  %{\smallaver{\Phi_{\vlambda,\vmu}^*| V_{\valpha}^*(1)|\Phi_i}_a
  %  \, \smallaver{\Phi_i |V_{\valpha}(1) |\Phi_{\vlambda,\vmu}}_b}
  \label{eq:Fi} \\
  \wh{\mathcal{F}}_{j,c,d}(z) &= \sum_n (z-1)^{-2h_{\valpha}+h_j^{(n)}}
  {\smallaver{\Phi_{\vlambda,\vmu}^*| \Phi_{\vlambda,\vmu}(1)|\wh\Phi_j^{(n)}}_c
    \, \smallaver{\wh\Phi_j^{(n)}|V_{\valpha}(1) |V_{\valpha}^*}_d}
  %{\smallaver{\Phi_{\vlambda,\vmu}^*| \Phi_{\vlambda,\vmu}(1)|\wh\Phi_j}_c
  %  \, \smallaver{\wh\Phi_j|V_{\valpha}(1) |V_{\valpha}^*}_d}
  \label{eq:Fj}  \,,
\end{align}
where the vectors $\ket{\Phi_i^{(m)}}$ (resp. $\ket{\wh\Phi_j^{(n)}}$) form an orthonormal basis of $W_3$ descendants of $\ket{\Phi_i}$ (resp. $\ket{\wh\Phi_j}$).
%, and the normalisation factors $A_i,\wh A_j$ are chosen so that $\mathcal{F}_i(z) \sim z^{-h_\Phi-h_{\valpha}+h_i}$ at $z \to 0$, and $\wh{\mathcal{F}}_j(z) \sim (z-1)^{-2h_{\valpha}+h_j}$ at $z \to 1$.
The indices $a,b,c,d$ in (\ref{eq:Fi}--\ref{eq:Fj}) denote the distinct possible structure constants involving a given $W_3$ descendant, in the case when the fusion of external fields produces internal fields with non-trivial multiplicities (see~\cite{BEFS16,BCES17}).
%For instance, the index $a$ takes values $1, 2, \dots \mathcal{N}_{\Phi,V_{\valpha}}^{\Phi_i}$.
The physical correlation function can be written in terms of conformal blocks:
\begin{equation} \label{eq:G-decomp}
  G(z,\zb) = \sum_{i,a,b} X_{i,a,b} \, |\mathcal{F}_{i,a,b}(z)|^2
  = \sum_{j,c,d} \wh{X}_{j,c,d} \, |\wh{\mathcal{F}}_{j,c,d}(z)|^2 \,,
\end{equation}
where the indices $a$ and $b$ take the values $1, 2, \dots \mathcal{N}_{\Phi,V_{\valpha}}^{\Phi_i}$, whereas $c=1,2 \dots \mathcal{N}_{\Phi,\Phi^*}^{\wh\Phi_j}$, and $d=1,2 \dots \mathcal{N}_{V_{\valpha}^{\phantom{*}},V_{\valpha}^*}^{\wh\Phi_j}$.

From the fusion rules~(\ref{eq:Phi.V}--\ref{eq:Phi.Phi}), one can describe the sets of possible internal primary fields $\Phi_i$ and $\wh\Phi_j$. Each $\Phi_i$ is of the form $V_{\valpha-b^{-1}\vlambda'+b\vmu'}$ with $\vlambda' \in [\vlambda]$ and $\vmu' \in [\vmu]$, and appears in~\eqref{eq:G-decomp} with multiplicity:
\begin{equation}
  (\mathcal{N}_{\Phi,V_{\valpha}}^{\Phi_i})^2= \left(\mathcal{N}_{\Phi_{\vlambda,\vmu},V_{\valpha}}^{V_{\valpha-b^{-1}\vlambda'+b\vmu'}}\right)^2
  = m_{\vlambda}^2(\vlambda')m_{\vmu}^2(\vmu') \,.
\end{equation}
Each $\wh\Phi_j$ is of the form $\wh\Phi_j = \Phi_{\wh{\vlambda},\wh{\vmu}}$,
where $[\wh{\vlambda}]$ (resp. $[\wh{\vmu}]$) is an $\sla_3$ irrep appearing in the fusion $[\vlambda] \times [\vlambda^*]$ (resp. $[\vmu] \times [\vmu^*]$).
Note that these internal representations are neutral: $q_{\wh\vlambda}=q_{\wh\vmu}=0$.
The field $\wh\Phi_j$ appears in~\eqref{eq:G-decomp} with multiplicity
\begin{equation}
  \mathcal{N}_{\Phi,\Phi^*}^{\wh\Phi_j} \times \mathcal{N}_{V_{\valpha},V_{\valpha}^*}^{\wh\Phi_j}
  = \mathcal{N}_{\Phi_{\vlambda,\vmu},\Phi_{\vlambda,\vmu}^*}^{\Phi_{\wh{\vlambda},\wh{\vmu}}}
  \times \mathcal{N}_{\Phi_{\wh{\vlambda},\wh{\vmu}},V_{\valpha}}^{V_{\valpha}}
  = N_{\vlambda \vlambda^*}^{\wh\vlambda} N_{\vmu \vmu^*}^{\wh\vmu} \times m_{\wh\vlambda}(0) m_{\wh\vmu}(0) \,.
\end{equation}
As a consistency check, let us compare the number of conformal blocks in the two channels:
\begin{align}
  \sum_i (\mathcal{N}_{\Phi,V_{\valpha}}^{\Phi_i})^2 &=
  \sum_{\vlambda' \in [\vlambda]}m_{\vlambda}^2(\vlambda') \times \sum_{\vmu' \in [\vmu]} m_{\vmu}^2(\vmu') \,, \\
  \sum_j \mathcal{N}_{\Phi,\Phi^*}^{\wh\Phi_j} \times \mathcal{N}_{V_{\valpha},V_{\valpha}^*}^{\wh\Phi_j}
  &= \sum_{[\wh\vlambda]} N_{\vlambda \vlambda^*}^{\wh\vlambda}\, m_{\wh\vlambda}(0) \times \sum_{[\wh\vmu]} N_{\vmu \vmu^*}^{\wh\vmu} \,  m_{\wh\vmu}(0) \,.
\end{align}
The two expressions coincide, because of the identity \eqref{eq:sum-m^2}.

\subsection{Rational models}

\begin{figure}[h]
  \begin{center}
    \includegraphics{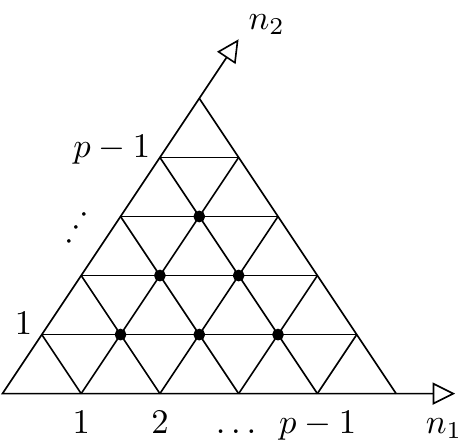}
    \qquad
    \includegraphics{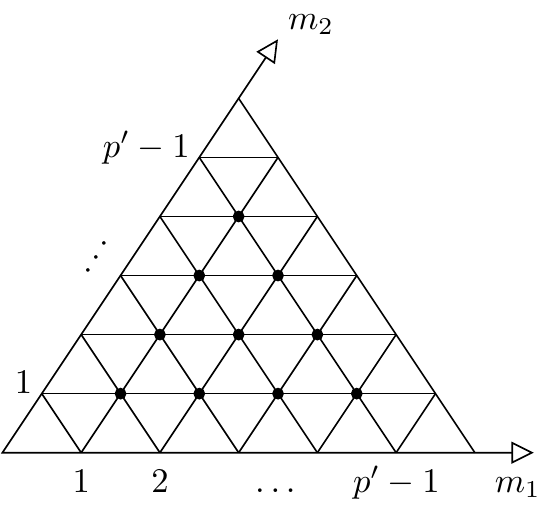}
    \caption{Possible values of Kac indices $(n_1,n_2)$ and $(m_1,m_2)$ for the rational model $\mathcal{M}_{p,p'}$. In this example, we have represented $\mathcal{M}_{4,5}$.}
    \label{fig:Kac-W3}
  \end{center}
\end{figure}

The rational model $\mathcal{M}_{p,p'}$ has $b=\sqrt{p/p'}$. Its operator algebra is finite, and consists of the fields in the $W_3$ Kac table \cite{FateevZamo87} :
\begin{equation}
  \mathcal{M}_{p,p'} = \left\{
  \Phi\kac{n_1 & m_1}{n_2 & m_2} \,, \quad n_1+n_2<p \,, \quad m_1+m_2<p'
  \right\} \,.
\end{equation}
This Kac table may be represented as the Cartesian product of two triangular tables: see \figref{Kac-W3}.
\bigskip

For any Kac indices $n_1,n_2,m_1,m_2$, and any real numbers $u,v$ one has:
\begin{equation}
  \valpha\kac{n_1 & m_1}{n_2 & m_2}
  = \valpha\kac{n_1+u p & m_1+u p'}{n_2+v p & m_2+v p'} \,.
\end{equation}
From the above relations, we get three sets of Kac indices for a given degenerate primary field (with equivalent vertex charges, related by Weyl rotations):
\begin{equation*}
  \Phi\kac{n_1 & m_1}{n_2 & m_2}
  \equiv \Phi\kac{p-n_1-n_2 & p'-m_1-m_2}{n_1 & m_1}
  \equiv \Phi\kac{n_2 & m_2}{p-n_1-n_2 & p'-m_1-m_2} \,.
\end{equation*}
The $\Zbb_3$ charge associated to this degenerate primary field is defined as:
\begin{equation} \label{eq:q}
  q \kac{n_1 & m_1}{n_2 & m_2} = \begin{cases}
    n_1-n_2 &\text{if } p \equiv 0 \mod 3 \\
    m_1-m_2 &\text{if } p' \equiv 0 \mod 3 \\
    (n_1-n_2)+(m_1-m_2) &\text{if } p+p'\equiv 0 \mod 3 \\
    (n_1-n_2)-(m_1-m_2) &\text{if } p-p'\equiv 0 \mod 3
  \end{cases}
\end{equation}
The four above cases are disjoint when $p,p'$ are coprime. The fusion rules for degenerate operators conserve this $\Zbb_3$ charge:
\begin{equation}
  \text{if} \quad \mathcal{N}_{\Phi,\Phi'}^{\Phi''} \neq 0 \,,
  \qquad \text{then} \quad q+q' \equiv q'' \mod 3 \,,
\end{equation}
where $\Phi,\Phi',\Phi''$ are degenerate fields of the form~\eqref{eq:Phi}, and $q,q',q''$ are the associated $\Zbb_3$ charges.
\bigskip

\subsection{Example: the three-state Potts model}

Let us describe in detail the operator content of the three-state Potts model, \textit{i.e.} the rational model $\mathcal{M}_{p,p'}$ with $(p,p')=(4,5)$ and central charge $c=4/5$ :
\begin{equation}
  \begin{aligned}
    \id &= \Phi\kac{1 & 1}{1 & 1} \equiv \Phi\kac{2 & 3}{1 & 1} \equiv \Phi\kac{1 & 1}{2 & 3} \,, \qquad q_\id= 0 \,, &&\qquad h_\id=0 \,, \\
  \sigma &= \Phi\kac{1 & 2}{1 & 1} \equiv \Phi\kac{2 & 2}{1 & 2} \equiv \Phi\kac{1 & 1}{2 & 2} \,,
    \qquad q_\sigma= +1 \,, &&\qquad  h_\sigma=\frac{1}{15} \,, \\
    \sigma^* &= \Phi\kac{1 & 1}{1 & 2} \equiv \Phi\kac{1 & 2}{2 & 2} \equiv \Phi\kac{2 & 2}{1 & 1} \,,
    \qquad  q_{\sigma^*}= -1 \,, &&\qquad h_{\sigma^*}=\frac{1}{15} \,, \\
    \psi &= \Phi\kac{1 & 1}{1 & 3} \equiv \Phi\kac{2 & 1}{1 & 1} \equiv \Phi\kac{1 & 3}{2 & 1} \,, \qquad q_\psi= +1 \,, &&\qquad h_\psi=\frac{2}{3} \,, \\
    \psi^* &= \Phi\kac{1 & 3}{1 & 1} \equiv \Phi\kac{1 & 1}{2 & 1} \equiv \Phi\kac{2 & 1}{1 & 3} \,, \qquad q_{\psi^*}= -1 \,, &&\qquad h_{\psi^*}=\frac{2}{3} \,, \\
    \eps &= \Phi\kac{1 & 2}{1 & 2} \equiv \Phi\kac{2 & 1}{1 & 2} \equiv \Phi\kac{1 & 2}{2 & 1} \,, \qquad q_\eps= 0 \,, &&\qquad h_\eps=\frac{2}{5} \,, \\
  \end{aligned}
\end{equation}
The fusion rules for the spin field $\sigma$ are:
\begin{equation}
  \sigma \times \sigma \to \sigma^* + \psi^* \,,
  \qquad \sigma \times \sigma^* \to \id + \eps \,.
\end{equation}
Hence, the spin field produces, by fusion with itself or its conjugate, the full $q=-1$ and $q=0$ sectors of the Kac table, respectively.

\section{Spin fields and their fusion rules}
\label{sec:spin}

\subsection{The Virasoro case}
\label{sec:Virasoro}

\paragraph{Coulomb-gas parameterisation.}
In the standard case of CFTs governed by the Virasoro algebra, the Coulomb-gas parameterisation for the central charge is \cite{DotsenkoFateev84}
\begin{equation} \label{eq:cc-Vir}
  c = 1 - 24Q^2 \,,
  \qquad Q = \frac{1}{2} \left(b^{-1} - b \right) \,,
\end{equation}
and the primary fields are represented by vertex operators $V_\alpha$, with conformal dimension
\begin{equation}
  \qquad h_\alpha = \alpha(\alpha-2Q) \,.
\end{equation}
Since $h_\alpha=h_{2Q-\alpha}$, one can identify $V_\alpha \equiv V_{2Q-\alpha}$. The degenerate fields have vertex charges and conformal dimensions of the form:
\begin{equation}
  \alpha_{rs} = \frac{1-r}{2} b^{-1} - \frac{1-s}{2} b \,,
  \qquad h_{rs} = \frac{(rb^{-1} - sb)^2 - (b^{-1}-b)^2}{4} \,,
\end{equation}
with $r,s$ positive integers. The corresponding field is denoted $\Phi_{rs}$.

\paragraph{Rational models.}
In rational models $\mathcal{M}_{p,p'}$, when $b=\sqrt{p/p'}$ and $p,p'$ are coprime integers, the degenerate field $\Phi_{rs}$ carries a $\Zbb_2$ charge given by
\begin{equation}
  q_{rs} = \begin{cases}
    r-1 & \text{if $p$ is even,} \\
    s-1 & \text{if $p'$ is even,} \\
    r+s & \text{if $p$ and $p'$ are odd.}
  \end{cases}
\end{equation}
The fusion rules between degenerate fields \cite{BPZ84} conserve this $\Zbb_2$ charge:
\begin{equation}
  \text{if} \quad \mathcal{N}_{\phi_{rs}, \phi_{r's'}}^{\phi_{r''s''}} \neq 0 \,,
  \qquad \text{then} \quad q_{rs}+q_{r's'} \equiv q_{r''s''} \mod 2 \,.
\end{equation}
In particular, the Ising model is given by the Virasoro rational model $\mathcal{M}_{3,4}$ with central charge $c=1/2$. It has three degenerate fields:
\begin{equation}
  \id = \Phi_{11}=\Phi_{23} \,,
  \qquad \eps = \Phi_{13}=\Phi_{21} \,,
  \qquad \sigma = \Phi_{12}=\Phi_{22} \,,
\end{equation}
with charges $q_{\id}=q_\eps=0$ and $q_\sigma=1$, and fusion rules:
\begin{equation}
  \sigma \times \sigma = \id + \eps \,,
  \qquad \eps \times \eps = \id \,.
\end{equation}
%\bigskip

\begin{figure}[h]
  \begin{center}
    \includegraphics{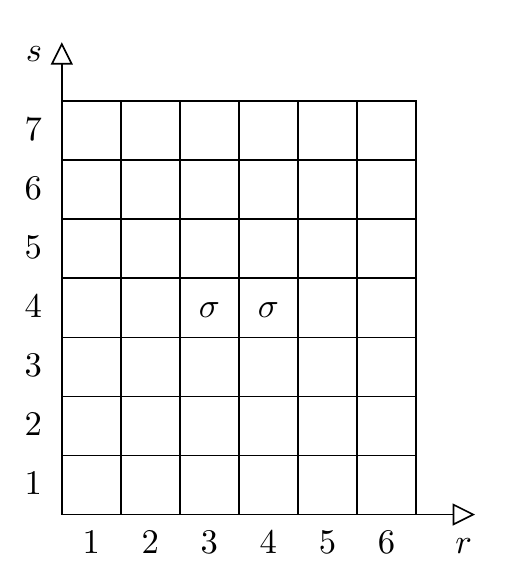}
    \caption{Position of the spin operator $\sigma$ in the Kac table of the Virasoro rational model $\mathcal{M}_{7,8}$.}
    \label{fig:Kac}
  \end{center}
\end{figure}

\paragraph{The $\Zbb_2$ spin field.}
For non-rational models, we introduce the generalised $\Zbb_2$ charge for degenerate fields (in analogy with the Ising model) $\wt{q}_{rs} = s-1$, and construct a generalisation of the $\Zbb_2$ spin field $\sigma$. 
The $\Zbb_2$-neutral sector of degenerate fields consists of the fields $\Phi_{rs}$ with $s$ odd. If we impose that the fusion of the spin field $\sigma$ with itself produces the full $\Zbb_2$-neutral sector:
\begin{equation} \label{eq:sigma.sigma->even}
  \sigma \times \sigma \to \sum_{\wt{q}_{rs} \equiv 0 \ [2]} \Phi_{rs} \,,
\end{equation}
then, by consistency of the operator algebra, the fusion $\Phi_{rs} \times \sigma \to \sigma$ with any even $\wt{q}_{rs}$ should be allowed. In particular, the fusion of $\Phi_{21}$ with a generic operator $V_\alpha$ is well known to be:
\begin{equation}
  \Phi_{21} \times V_\alpha \to V_{\alpha+b^{-1}/2} + V_{\alpha-b^{-1}/2} \,.
\end{equation}
If impose $V_{\alpha} \equiv V_{\alpha \pm b^{-1}/2}$, we get $\alpha=\alpha_{\pm 1/2,0}$. Hence, the only field possibly consistent with~\eqref{eq:sigma.sigma->even} is
\begin{equation} \label{eq:spin-Z2}
  \sigma = \Phi_{1/2,0} \equiv \Phi_{-1/2,0} \,.
\end{equation}
Repeating this argument with the fusion rules between degenerate and generic fields:
\begin{equation}
  \Phi_{rs} \times V_\alpha = \sum_{j=0}^{r-1} \sum_{k=0}^{s-1} V_{\alpha + (\frac{1-r}{2}+j)b^{-1} - (\frac{1-s}{2}+k)b} \,,
\end{equation}
one can easily see that, with the choice~\eqref{eq:spin-Z2}, all $\Zbb_2$-neutral degenerate fields are also allowed in the fusion \eqref{eq:sigma.sigma->even}. For the rational models $\mathcal{M}_{p,p'}$ with $p$ odd and $p'$ even, the operator \eqref{eq:spin-Z2} becomes degenerate:
\begin{equation}
  \sigma \equiv \Phi_{\frac{p-1}{2},\frac{p'}{2}} \equiv \Phi_{\frac{p+1}{2},\frac{p'}{2}} \,,
\end{equation}
and sits at the center of the Kac table: see \figref{Kac}.
One can easily show that, like in non-rational models, the fusion $\sigma \times \sigma$ produces the full $\wt{q}=0$ sector of the Kac table. For the Ising model, one recovers the spin operator $\sigma=\Phi_{12}$.

\paragraph{Potts and O($n$) critical lines.} Remarkably, the field~\eqref{eq:spin-Z2} has a geometrical interpretation for the two well-known continuous families of CFTs with Virasoro symmetry: the critical Potts and O($n$) models.

The $\mathcal{Q}$-state Potts model admits a cluster expansion, the Fortuin-Kasteleyn (FK) model, where $\mathcal{Q}$ becomes the fugacity of a connected component (cluster). The critical line $0<\mathcal{Q} \leq 4$ is described by a CFT with parameter $b$ in \eqref{eq:cc-Vir} given by \cite{Nienhuis84}:
% \begin{equation}
%   \sqrt{\mathcal{Q}} = -2 \cos \pi b^2 \,, \qquad \frac{1}{\sqrt 2}<b \leq 1 \,.
% \end{equation}
\begin{equation}\label{eq:tric-crit-Potts}
  \sqrt{\mathcal{Q}} = -2 \cos \pi b^2 \,, \qquad \begin{cases}
    1 \leq b < {\sqrt \frac{3}{2}} & \text{for the tricritical point,} \\
    \frac{1}{\sqrt 2}<b \leq 1  & \text{for the critical point.}
  \end{cases}
\end{equation}

The field~\eqref{eq:spin-Z2} is exactly the FK spin operator, {\it i.e.} any correlation function with $\sigma$ operators discards the cluster configurations where a connected component contains a single $\sigma$. The energy operator of the Potts model is $\varepsilon=\Phi_{21}$ \cite{DotsenkoFateev84}.
%For integer $\mathcal{Q}$, the field $\sigma$ corresponds to the Potts spin operator.

In the O($n$) loop model, the parameterisation of the loop fugacity $-2 < n \leq 2$ takes the form:
\begin{equation}\label{eq:dense-dilute-On}
  n = -2 \cos \frac{\pi}{b^2} \,, \qquad \begin{cases}
    1 \leq b < \infty & \text{for the dense phase,} \\
    \frac{1}{\sqrt 2} \leq b \leq 1 & \text{for the dilute critical point.}
  \end{cases}
\end{equation}
In this model, the field~\eqref{eq:spin-Z2} corresponds to the ``one-leg'' operator, {\it i.e.} the operator inserting the end of an open path \cite{Nienhuis84}. The energy operator of the O($n$) model is $\varepsilon=\Phi_{13}$ \cite{DotsenkoFateev84}.

\begin{figure}[h]
  \begin{center}
    \includegraphics[width=0.8\linewidth]{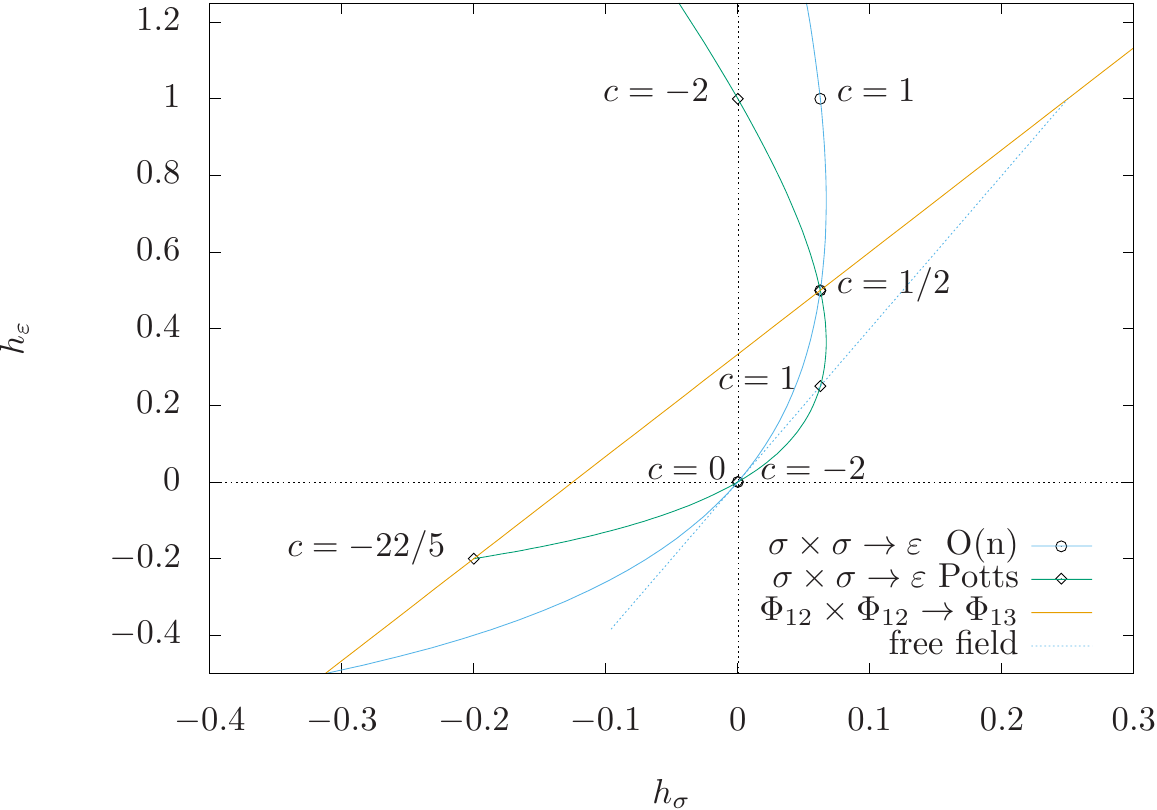}
    \caption{Conformal dimensions in the Virasoro fusion $\sigma \times \sigma \to \varepsilon$ for the O($n$) model and that for the Potts model. The central charge $c$ is labeled on the right (resp. left) for the O($n$) (resp. Potts) curve. The line for the fusion $\Phi_{12}\times \Phi_{12}\to \Phi_{13}$ is also plotted. The intersection $c=1/2$ at $(h_\sigma,h_\varepsilon)=(1/16,1/2)$ corresponds to the kink of the unitarity bound for the $\Zbb_2$ sum rule.}
    \label{fig:virplot}
  \end{center}
\end{figure}
In \figref{virplot}, we show three curves of the conformal dimensions $(h_\sigma,h_\varepsilon)$ for the the O($n$) model and that for the Potts model as well as $(h_{12}, h_{13})$
 in the range $1/2\leq b\leq \sqrt{5/2}$ , which contains parts not in \eqref{eq:tric-crit-Potts} and \eqref{eq:dense-dilute-On}.
This extension allows the curves to enter the third quadrant.
A few remarks are in order about this diagram, which are deeply related to the conformal bootstrap.
Many of these properties on the Virasoro CFTs are generalised to the $W_3$ CFTs (see \secref{phasediagram}): %%%%%%%%%%
\begin{itemize}
\item The unitarity bound \cite{Ising3d} obtained from the $\Zbb_2$ sum rule using the single correlation function $\langle \sigma\sigma\sigma\sigma \rangle$ of the lowest scaling dimensions 
 is close to the portion of the O($n$) curve connecting $c=-2$ to $c=1/2$ of the kink [the Ising model $(h_\sigma,h_\varepsilon)=(1/16,1/2)$ is realised at both $n=1$ and $\mathcal{Q}=2$],  %$(h_\sigma,h_\varepsilon)=(1/16,1/2)$
continued by the half line representing the fusion $\Phi_{12}\times \Phi_{12}\to \Phi_{13}$ for $h_{12}>1/16$. 
\item More concretely, the numerical bound \cite{Ising3d} goes slightly below (resp. above) the O($n$) curve
$h_\sigma=\frac{2 h_{\varepsilon }-h_{\varepsilon }^2}{8 \left(h_{\varepsilon }+1\right)}$ 
(resp. the half line $h_{13}=\frac{8h_{12}+1}{3}$) on the left (resp. right) of the kink.
Both the O($n$) curve ($n=c=-2$) and the unitarity bound starts with the slope $4$ at $(0, 0)$.
This reflects the fact that the fusion reduces to that of free field type (see the last remark on \figref{plot2}).
\item The half line of the fusion $\Phi_{12}\times \Phi_{12}\to \Phi_{13}$ is realised in the OPE $\varepsilon\times \varepsilon = \id  + \varepsilon'$ 
with $\varepsilon=\Phi_{12}$ and $\varepsilon'=\Phi_{13}$ in the thermal subsector of 
the tricritical $\mathcal{Q}$-state Potts model\footnote{The tricritical point with $\mathcal{Q}=2$ is the Virasoro rational model $\mathcal{M}_{4,5}$. Along this half line, aiming at understanding the numerical unitarity bounds, it was conjectured \cite{Liendo13bootstrapboundary} and has recently been proved \cite{Behan18unitarysubsector} that all the $\sla_2$ (global) conformal blocks appear with positive coefficients in the correlation function $\aver{\Phi_{12}\Phi_{12}\Phi_{12}\Phi_{12}}$.}
%only two Virasoro conformal blocks (from the channels $\id$ and $\varepsilon'$)
for generic $\mathcal{Q}$.
The analogue of $\Phi_{12}$ in the $W_3$ case is identified as the field $\sigma''$ in \eqref{eq:sigma''}.
\item The Potts curve $h_{\sigma}=\frac{h_{\varepsilon }-h_{\varepsilon }^2}{2 \left(2 h_{\varepsilon }+1\right)}$ 
in the first quadrant describes the dimensions $(h_\sigma,h_\varepsilon)$ for both the critical point and tricritical point.
These two series merge at $(1/16,1/4)$ with $\mathcal{Q}=4$ ($b=1$ and $c=1$).
If one further continues the tricritical (i.e. lower branch from $\mathcal{Q}=0$ at $(0,0)$ with $c=0$, it intersects at $h_\sigma=h_{\varepsilon}=-1/5$ with the other half of the thermal-subsector line ($b=\sqrt{5/2}$ and $c=-22/5$ along the Potts curve). 
This point also corresponds to the Virasoro rational model $\mathcal{M}_{2,5}$ of $c=-22/5$ (the Lee-Yang model realized at $b=\sqrt{2/5}$) characterized by
the only one primary operator of the dimension  $h_{12}=h_{13}=-1/5$.
The analogue of this highly degenerate point in the $W_3$ case has $c=-10$ (See \figref{plot}).
\item The O($n$) model at $n=2$ and the Potts model at $\mathcal{Q}=4$ have $(c, h_\sigma)=(1, 1/16)$. For this particular combination, the conformal block appearing in the 4-point function $\aver{\sigma\sigma\sigma\sigma}$ 
has a closed form for any values of the conformal dimensions of intermediate channels \cite{Zamolodchikov86_c1_exact_4pt}.
%%%%%%%%%%%%%%%%%%%%%%%%%%%%%%%%%%%%%%%%%%%%%%%%%%%
%%%%%%%%%%%%%%%%%%%%%%%%%%%%%%%%%%%%%%%%%%%%%%%%%%%
%%%%%%%%%%%%%%%%%%%%%%%%%%%%%%%%%%%%%%%%%%%%%%%%%%%
%%%%%%%%%%%%%%%%%%%%%%%%%%%%%%%%%%%%%%%%%%%%%%%%%%%
\end{itemize}
\subsection{Properties of a spin field in $W_3$ CFTs}
\label{sec:prop}

We consider the critical line of $W_3$ CFTs, with $0<b<\infty$ in (\ref{eq:Q}--\ref{eq:cc}). We shall identify some primary ``spin field'' $\sigma$, with the following required properties:
\begin{enumerate}
\item At $b=\sqrt{4/5}$, the field $\sigma$ should coincide with the spin field of the three-state Potts model.
%\item For rational values of $b^2=p/p'$, the field $\sigma$ should be  degenerate, and have a $\Zbb_3$ charge $q_\sigma=1$ when $p+p' \equiv 0 \mod 3$.
\item For any generic value of $b$, the field $\sigma$ should be parameterised in~\eqref{eq:kac} with rational Kac indices, independent of $b$, and the associated $\Zbb_3$ charge should be $\wt q_\sigma=1$, where $\wt q$ is defined as
  \begin{equation}
    \wt{q}\kac{n_1 & m_1}{n_2 & m_2} = (n_1-n_2)+(m_1-m_2) \,,
  \end{equation}
\textit{i.e.} it is the generalisation of the three-state Potts model's $\Zbb_3$ charge to a CFT with generic $b$ [see \eqref{eq:q}].
\item For any generic value of $b$, the fusions $\sigma \times \sigma$ and $\sigma \times \sigma^*$ should produce as many degenerate fields as possible, respectively in the $\wt q=-1$ and $\wt q=0$ sectors.
\end{enumerate}

\subsection{Fusion of a spin field with itself}
\label{sec:sigma.sigma}

Let us represent the spin field as a vertex operator $\sigma=V_{\valpha}$. Moreover, let us consider a degenerate operator $\Phi_{\vlambda,\vmu}$ associated to the pair of irreps $([\vlambda],[\vmu])$, as in \eqref{eq:Phi}.
By consistency of the operator algebra, one has the following equivalence:
\begin{equation} \label{eq:V.V}
  V_{\valpha} \times V_{\valpha} \to \Phi^*_{\vlambda,\vmu}+ \dots
  \qquad \Leftrightarrow \qquad
  \Phi_{\vlambda,\vmu} \times V_{\valpha} \to V_{\valpha}^* +\dots
\end{equation}
Using the fusion rule~\eqref{eq:Phi.V}, the above fusion is allowed iff $\valpha^*$ is equivalent [modulo the Weyl group action \eqref{eq:x.alpha}] to a charge in the right-hand-side of~\eqref{eq:Phi.V}
\begin{equation} \label{eq:constraint}
  \valpha^* \equiv \valpha-b^{-1}\vlambda'+b\vmu' \,,
\end{equation}
where $\vlambda' \in [\vlambda]$ and $\vmu' \in [\vmu]$.
If we impose that the fusion of $\sigma$ with itself produces the full $\wt q=-1$ sector, {\textit i.e.}
\begin{equation}
  \sigma \times \sigma \to \sum_{q_{\vlambda}+q_{\vmu} \equiv 1 \, [3]} \Phi^*_{\vlambda,\vmu} + \dots
\end{equation}
then the unique solution satisfying the properties of \secref{prop} is
\begin{equation} \label{eq:sigma}
  \sigma = \Phi\kac{\frac{2}{3} & \frac{1}{3}}{-\frac{1}{3} & \frac{1}{3}} \,,
\end{equation}
with eigenvalues:
\begin{equation}
  h_\sigma = \frac{1-8(b^{-1}-b)^2}{9} \,,
  \qquad w_\sigma = -\frac{2\beta\sqrt{3}}{27}(b^{-1}-b) \,.
\end{equation}
This can be proven by first imposing the constraint~\eqref{eq:constraint} for
$$(\vlambda,\vmu) \in \{ \quad ([\om_1],0), \quad (0,[\om_1]), \quad ([\om_2],[\om_2]) \quad \} \,,$$
and then using the fact that the weights of these representations are actually included in all representations with the same $\Zbb_3$ charges $(q_{\vlambda},q_{\vmu})$.

For rational values of $b^2=p/p'$ with $p+p' \equiv 0 \mod 3$, one can write:
\begin{align}
  &\sigma \equiv \Phi\kac{\frac{p+2}{3} & \frac{p'+1}{3}}{\frac{p-1}{3} & \frac{p'+1}{3}}
  \qquad \text{for} \quad (p,p') \equiv (+1,-1) \mod 3 \,, \\
  &\sigma \equiv \Phi\kac{\frac{p+1}{3} & \frac{p'-1}{3}}{\frac{p-2}{3} & \frac{p'-1}{3}} 
  \qquad \text{for} \quad (p,p') \equiv (-1,+1) \mod 3 \,,
\end{align}
so that the Kac indices are positive integers, and hence $\sigma$ is degenerate.
Note that the corresponding indices sit as close as possible to the center of the Kac table (see \figref{Kac-W3}).
Nicely, this simple rule for locating the spin field in $W_3$ CFTs turns out to be a direct generalization of that in the Virasoro CFTs (see also \figref{Kac}). %HS

\bigskip

If we impose a weaker condition on the right-hand side of the above fusion, namely
\begin{equation}
  \sigma' \times \sigma' \to \sum_{q_{\vlambda} \equiv 1 \,, \ q_{\vmu} \equiv 0 \ [3]}
  \Phi^*_{\vlambda,\vmu} + \dots
\end{equation}
we find an additional solution:
\begin{equation}
  \sigma' = \Phi\kac{1 & \frac{1}{2}}{0 & \frac{1}{2}} \,,
\end{equation}
with eigenvalues
\begin{equation}
  h_{\sigma'} = \frac{b^{-2}}{12} - \frac{3}{4}(b^{-1}-b)^2 \,,
  \qquad w_{\sigma'} = \frac{\beta\sqrt{3}}{54}b^{-1}(2b^{-1}-3b)(4b^{-1}-3b) \,.
\end{equation}
For generic $b$, the field $\sigma'$ is semi-degenerate at level one.
\bigskip

If we relax the third condition of \secref{prop}, but impose that the spin field be degenerate for any value of $b$ and obey the fusion rule:
\begin{equation}
  \sigma'' \times \sigma'' \to (\sigma'')^* + \Phi^* \kac{1 & 1}{1 & 3} + \dots
\end{equation}
we find the field
\begin{equation}\label{eq:sigma''}
  \sigma'' = \Phi\kac{1 & 2}{1 & 1} \,,
\end{equation}
with eigenvalues
\begin{equation}
  h_{\sigma''} = \frac{1}{3}(4b^2-3) \,,
  \qquad w_{\sigma''} = -\frac{2\beta\sqrt{3}}{27}b(3b^{-1}-4b)(3b^{-1}-5b) \,.
\end{equation}

\begin{figure}[h]
  \begin{center}
    \includegraphics[width=0.7\linewidth]{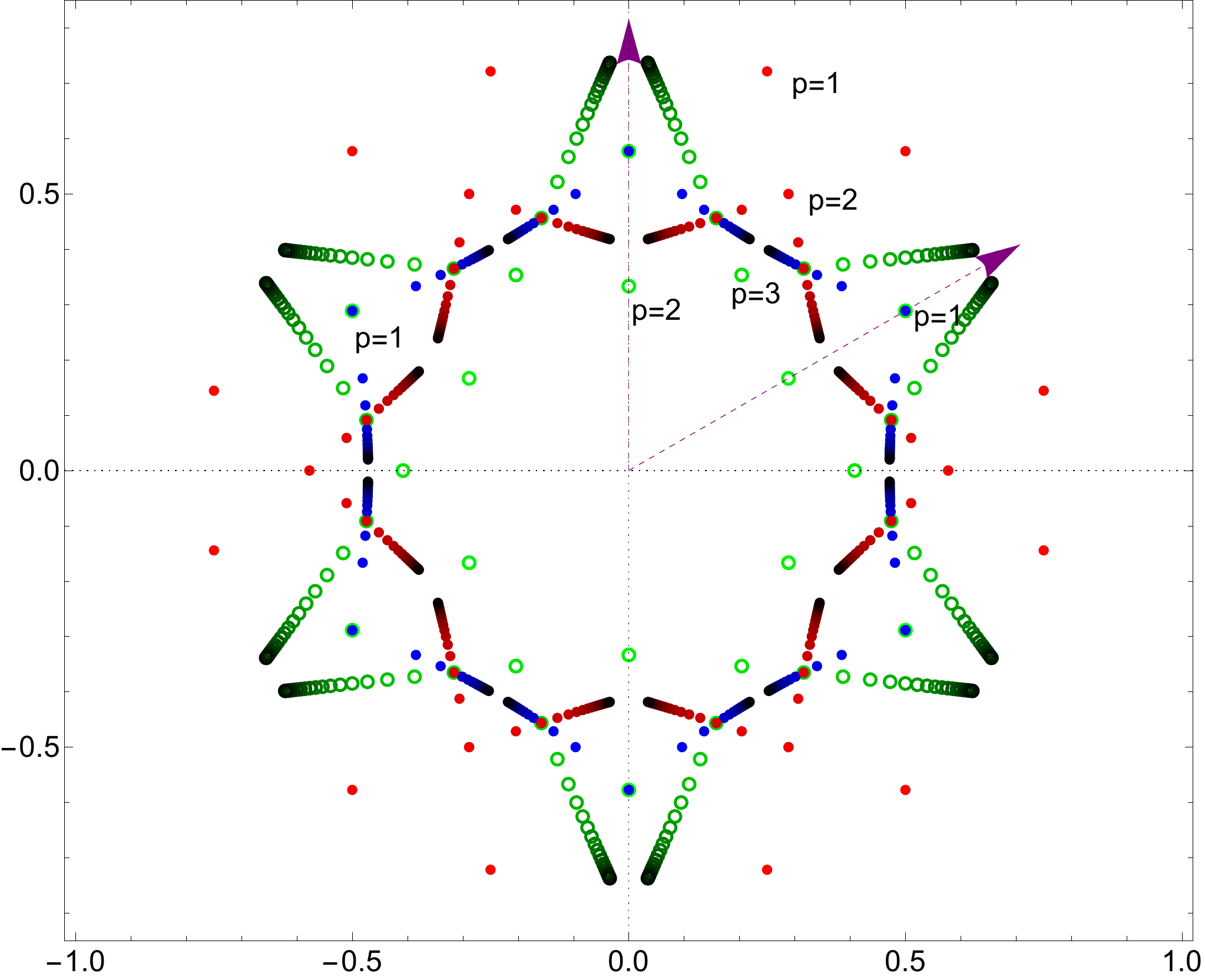}
    \caption{The orbits of 12 equivalent vertex charges $\valpha-\vQ$ for the fundamental spin field $\sigma$ (blue) and those for its generalizations $\sigma'$ (red) and $\sigma''$ (green open circle) as the central charge increased from 
$c=-10$ ($p=1$) to $c=2$ ($p=\infty$) with $b^2=p/(p+1)$. 
The markers for integers $p\in [1, 20]$ are plotted for each orbit, which gets darker as $p$ increases.
Each orbit is along a quadratic curve, which looks almost linear at this scale. 
}
    \label{fig:12WeylOrbits}
  \end{center}
\end{figure}

We now comment on the geometrical properties of $\sigma$, $\sigma'$, and $\sigma''$ on the plane of the (shifted) vertex charges $\valpha-\vQ$ [see \eqref{eq:Q} and \eqref{eq:kac}]. 
In \figref{12WeylOrbits}, we plot the orbits of all the vertex charges $\valpha-\vQ$ (equivalent under the Weyl group) of these fields as well as their $w$-charge-conjugates $\valpha^*-\vQ$ ($\mathbb{Z}_3$-charge conjugates: $\sigma^{*}$, $(\sigma')^{*}$, $(\sigma'')^{*}$)  realised for the central charge $-10\leq c\leq 2$ ($p\geq 1$) with $b^2=p/(p+1)$. 
See \eqref{eq:x.alpha}, \eqref{eq:alpha.star}, and Appendix \ref{sec:sl3} for how to move the vertex charges under the Weyl group and the conjugation.
Here we take $p'=p+1$ and show only the case with $b\leq 1$.
The fundamental weights \eqref{eq:weightsroots}, shown in purple, are set as $\om_1=(\sqrt{3},1)/\sqrt{6}$ and $\om_2=(0, \sqrt{2/3})$. Among the six directions from the origin, the three directions towards the representation $[\om_1]$ in \eqref{eq:fundamentalrep} are for charge $\tilde{q}=1$, and the rest towards $[\om_2]$ are for $\tilde{q}=-1$.

Note that there are $12 = 2$ (charge conjugations) %$\times$ $2$ ($w$-conjugations) 
$\times$ 6 (Weyl group actions) orbits of vertex charges that yield the same conformal dimension.
Using this $12$-fold symmetry, one could fold the obits into a narrow fan region (of $\pi/6<\arg (\valpha-\vQ)<\pi/3$, for instance).
Then the orbit of $\sigma'$ and that of $\sigma''$ respectively needs to be reflected one and three times at the walls of the fan, while that of the fundamental spin field $\sigma$ simply moves from one wall to the opposite wall.
By construction (the first condition of \secref{prop}), three orbits intersect simultaneously at $c=4/5$ ($p=4$).
In addition, the orbit of $\sigma$ and that of $\sigma''$ intersect on the wall at $c=-10$ ($p=1$). 
In the other branch $b^2>1$ ($p<-1$), the traces of $\sigma$ remain the same, while those of $\sigma'$ and $\sigma''$ change their shapes; 
at $c=-10$ ($p=-2$, $b^2=2$), $\sigma$ coincides on the wall with $\sigma'$ instead of $\sigma''$. This point is revisited in \secref{phasediagram}. 

%Once one identifies several important fields in the continuous family of CFTs, this diagram can be used to single out interesting models 
On both branches, we may observe a nice geometrical property of the three fields if we focus on the fundamental right triangle formed by $\om_1$ and $\om_2$.
Namely, as $c\to 2$, the fundamental spin field $\sigma$ tends to the center of {\it its face} (see \eqref{eq:A2} for the role of this field at $c=2$); analogously, $\sigma'$ tends to the midpoint of {\it its edge} and $\sigma''$ tends to {\it its vertex}.

%%%%%%%%%%%%%%%%%%%%%%%%%%%%%%%%%%%%%%%%%%%%%%%%%%%
%%%%%%%%%%%%%%%%%%%%%%%%%%%%%%%%%%%%%%%%%%%%%%%%%%%
\subsection{Fusion of a spin field with its conjugate}
\label{sec:sigma.sigma*}

To study the possible degenerate fields produced by the fusion $\sigma \times \sigma^*$, we use the equivalence analogous to~\eqref{eq:V.V}:
\begin{equation} \label{eq:V.V*}
  V_{\valpha} \times V_{\valpha}^* \to \Phi_{\vlambda,\vmu}+ \dots
  \qquad \Leftrightarrow \qquad
  \Phi_{\vlambda,\vmu} \times V_{\valpha} \to V_{\valpha} +\dots
\end{equation}
Hence the presence of a given completely degenerate field in the fusion $\sigma \times \sigma^*$ can be determined by using again \eqref{eq:Phi.V}. A similar approach can be used for the fields $\sigma'$ and $\sigma''$. Reasoning as in~\secref{sigma.sigma}, one obtains:
\begin{align}
  \sigma \times \sigma^* &\to
  \sum_{q_{\vlambda}+q_{\vmu} \equiv 0 \ [3]} \Phi_{\vlambda,\vmu} + \dots \\
  \sigma' \times (\sigma')^* &\to \sum_{q_{\vlambda} \equiv 0 \,, \ q_{\vmu} \equiv 0 \ [3]} \Phi_{\vlambda,\vmu} + \dots  \\
  \sigma'' \times (\sigma'')^* &\to \id + \Phi\kac{1 & 2}{1 & 2} \,,
\end{align}
where the dots denote primary fields which are not completely degenerate.

\subsection{Phase diagrams of fusion processes}
\label{sec:phasediagram}

\begin{figure}[h]
  \begin{center}
    \includegraphics[width=0.8\linewidth]{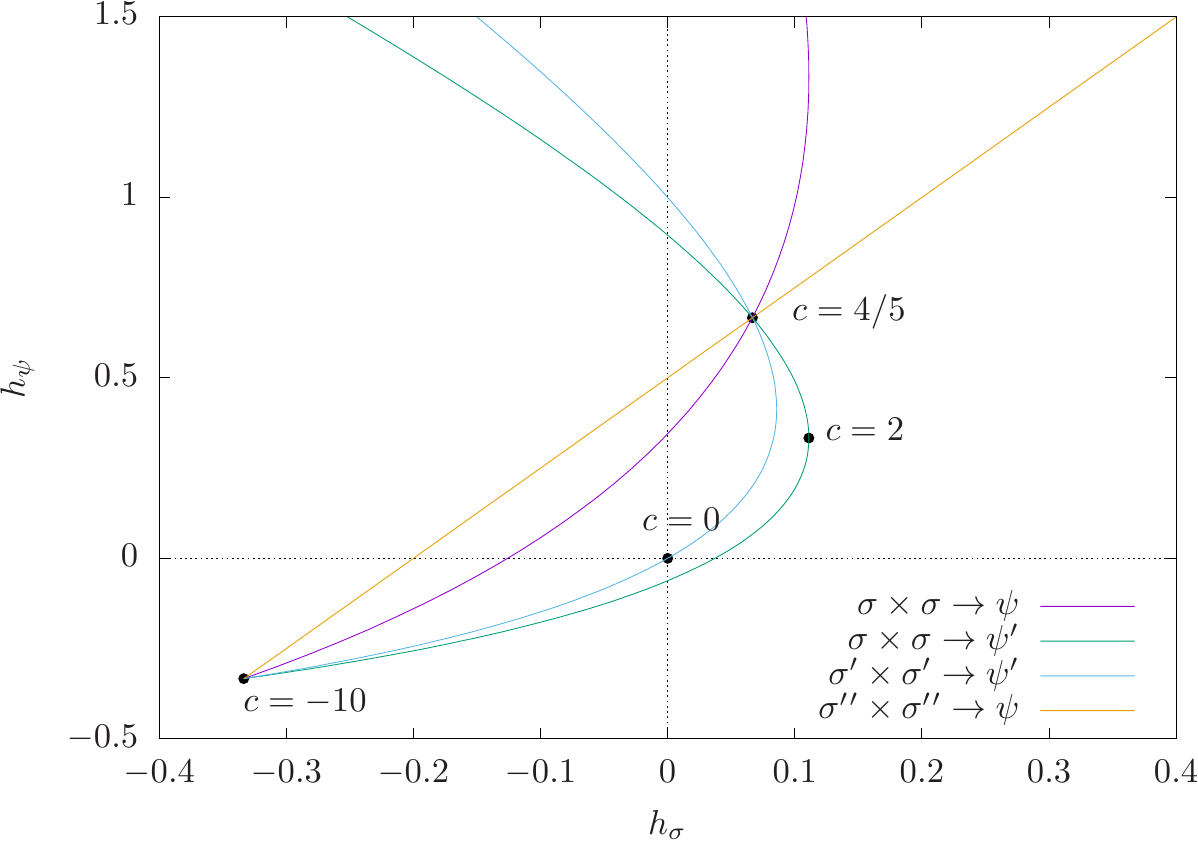}
    \caption{Phase diagram for the fusion $\sigma \times \sigma \to \psi^*$ and its variants.}
    \label{fig:plot}
  \end{center}
\end{figure}

For later convenience, we introduce the notations:
\begin{equation}
  \psi = \Phi\kac{1 & 1}{1 & 3} \,,
  \qquad \psi' = \Phi\kac{2 & 1}{1 & 1} \,,
  \qquad \eps = \Phi\kac{1 & 2}{1 & 2} \,.
\end{equation}
From the above discussion, we have the fusion rules:
\begin{equation}
  \begin{aligned}
    &\sigma \times \sigma \to \psi^* + (\psi')^* + \dots
    &\qquad\qquad& \sigma \times \sigma^* \to \id + \eps + \dots \\
    &\sigma' \times \sigma' \to (\psi')^* + \dots
    && \sigma' \times (\sigma')^* \to \id + \eps + \dots \\
    &\sigma'' \times \sigma'' \to (\sigma'')^* + \psi^*
    && \sigma'' \times (\sigma'')^* \to \id + \eps \,.
  \end{aligned}
\end{equation}

In \figref{plot}, we show the conformal dimensions $(h_\sigma,h_\psi), (h_\sigma,h_{\psi'}), (h_\sigma',h_{\psi'}),$ and $(h_\sigma'',h_{\psi})$ in the range $1/\sqrt{2}<b<\sqrt{2}$. %\dots replaced. typo corrected:  1/\sqrt -> 1/\sqrt{2}
Let us comment on some special points on these curves:
\begin{itemize}
\item At central charge $c=4/5$ ($b=\sqrt{4/5}$), we recover the three-state Potts model, and the various operators coincide:
  $$h_\sigma = h_{\sigma'} = h_{\sigma''} = \frac{1}{15} \,, \qquad h_\psi = h_{\psi'}=\frac{2}{3} \,.$$
\item At central charge $c=2$, we have $h_\sigma=1/9$ and $h_\psi=1/3$. Let us compare this to the $b \to 1$ limit of the CG action, which admits more screening charges than for generic $b$ :
  \begin{equation} \label{eq:A2}
  \mathcal{A}[\vect\phi] \to \int \frac{d^2x}{8\pi} \sqrt{|g|} \left[
    \partial_\mu\vect\phi \cdot \partial^\mu\vect\phi
    + \sum_{\ve \in \{\pm \ve_1, \pm \ve_2, \pm\vect\rho\}} \normal{e^{i\ve \cdot \vect\phi}}
    \right] \,.
\end{equation}
  This action is compatible with a compactification condition $\vect\phi \equiv \vect\phi+2\pi \mathcal{R}$ and the $\mathbb{Z}_3$ symmetry $\vect\phi \equiv \theta.\vect\phi$, where $\theta$ is the rotation of angle $\frac{2\pi}{3}$ in the $\vect\phi$ plane: thus, with these identifications, the sector without screening charges is identical to the $\mathbb{Z}_3$ orbifold of the complex boson \cite{DFMS87}. In this orbifold theory, the twist field has conformal dimension $1/9$, which we identify as the field $\sigma$.
\item  It would be natural to consider that the above combination $(c, h_\sigma)=(2, 1/9)$ would generalize $(c, h_\sigma)=(1, 1/16)$ in the Virasoro case \cite{Zamolodchikov86_c1_exact_4pt}, where one has the conformal block in a closed form.
            In order to analyse the exponential decay of the OPE coefficients for higher conformal dimension operators, this point would become an important reference point.  
%%
%%Could be moved to the Conclusion and Perspective section
%%
\item At central charge $c=0$ ($b=\sqrt{4/3}$), we get $h_{\sigma'}=h_{\psi'}=0$.
\item At central charge $c=-10$, if we choose $b=1/\sqrt{2}$ (resp. $b=\sqrt{2}$) we have $h_\sigma=h_{\sigma''}=h_\psi=-1/3$ (resp. $h_\sigma=h_{\sigma'}=h_{\psi'}=-1/3$).
\end{itemize}

\begin{figure}[h]
  \begin{center}
    \includegraphics[width=0.8\linewidth]{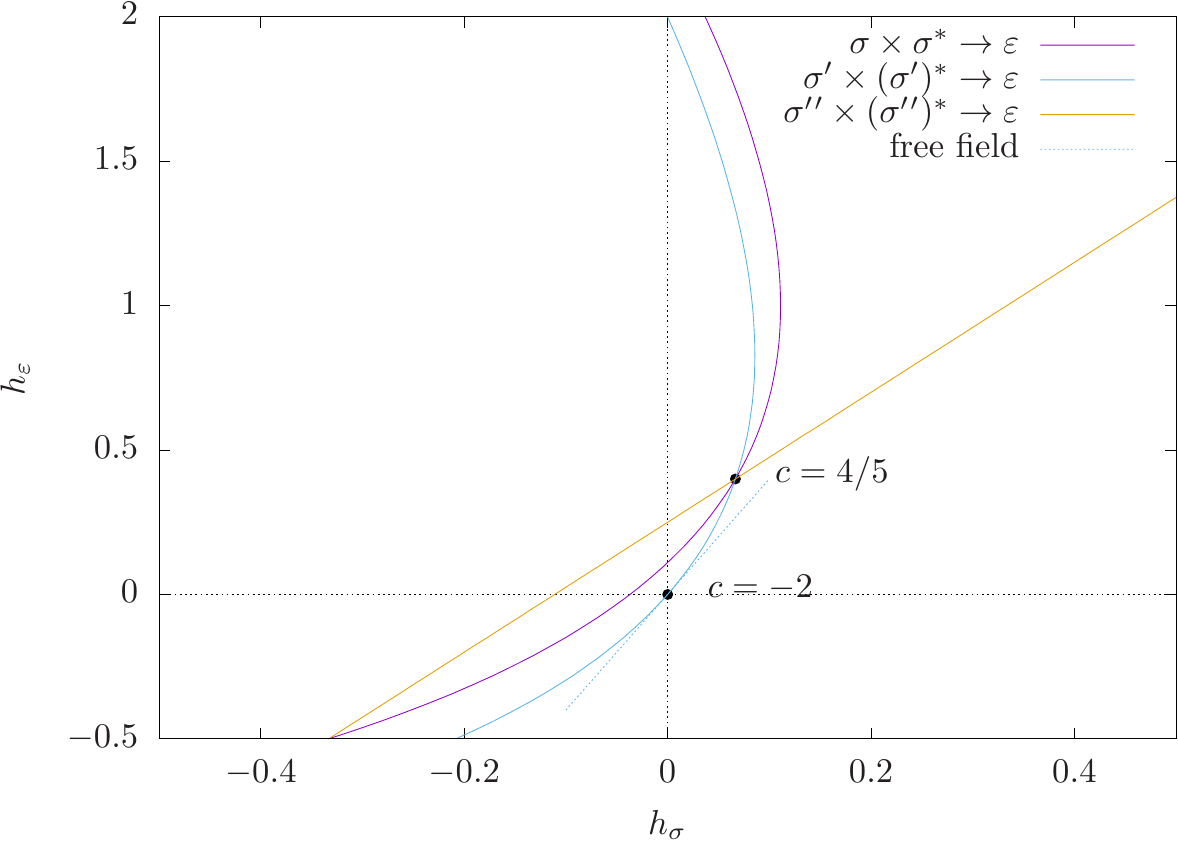}
    \caption{Phase diagram for the fusion $\sigma \times \sigma^* \to \eps$ and its variants.}
    \label{fig:plot2}
  \end{center}
\end{figure}

In \figref{plot2}, we show the conformal dimensions $(h_\sigma,h_\eps), (h_{\sigma'},h_\eps)$, and $(h_{\sigma''},h_\eps)$. The two points of interest are:
\begin{itemize}
\item At the three-state Potts point, we recover $h_\sigma=1/15, h_\eps=2/5$.
\item At central charge $c=-2$ ($b=\sqrt{2/3}$) we have $h_{\sigma'}=h_\eps=0$. In the vicinity of this point, the relation $h_\eps \sim 4h_{\sigma'}$ holds. This is compatible with a free-field action, where the vertex operator $\sigma \sim V_{\valpha}$ has dimension $h_{\valpha}=\valpha^2/2$, and the fusion rules are trivial: $V_{\valpha} \times V_{\valpha} \to V_{2\valpha}$.
\end{itemize}

\section{Conclusion and perspectives}
\label{sec:concl}

By studying carefully the fusion rules in generic $W_3$ CFTs, we have identified the fundamental spin field $\sigma$ \eqref{eq:sigma}, whose fusion with itself and its conjugate generates as many primary fields as allowed by the internal $\Zbb_3$ symmetry of the fusion rules. Some variants $\sigma'$ and $\sigma''$ are also 
obtained by relaxing the conditions on the fusion. %%%
The vertex charges of the fields $\sigma$, $\sigma'$, and $\sigma''$ coincide at the three-state Potts model ($c=4/5$), and as $c\to 2$, 
tend to the face center, the midpoint of the edge, and the vertex of the fundamental triangle, respectively. %formed by $\om_1$ and $\om_2$

The present results may serve as a basis to apply a conformal bootstrap approach, {\it i.e.} to determine the regions of the phase diagrams $(h_\sigma,h_\psi)$ and  $(h_\sigma,h_\eps)$ where the structure constants involved in the four-point function $\aver{\sigma\sigma\sigma\sigma}$ are positive. Note that some numerical bootstrap results related to $W_3$ CFTs are reported in \cite{RongSu18}. One may expect that, like in the O($n$) model for %%%
the Virasoro case, some of the exact curves in \figref{plot} and \figref{plot2} give good approximations to the boundaries of these regions.
In order to deepen our analytic understanding, it would also be useful to quantitatively study the infinite OPEs between the fundamental spin fields at generic points in the one-parameter family of $W_3$ CFTs and to see the exact pattern of the weak unitarity violation.

%%%%%%%%%%%%%%%%%%%%%%%%%%%%%%%%%%%%%%%%%%%%%%%%%%%%%%%%%%%%%%%%%%%%%%%%%%%%%%%
%%%%%%%%%%%%%%%%%%%%%%%%%%%%%%%%%%%%%%%%%%%%%%%%%%%%%%%%%%%%%%%%%%%%%%%%%%%%%%%
%%%%%%%%%%%%%%%%%%%%%%%%%%%%%%%%%%%%%%%%%%%%%%%%%%%%%%%%%%%%%%%%%%%%%%%%%%%%%%%
%%%%%%%%%%%%%%%%%%%%%%%%%%%%%%%%%%%%%%%%%%%%%%%%%%%%%%%%%%%%%%%%%%%%%%%%%%%%%%%

\section*{Appendix}
\appendix

\section{Representation theory of the $\sla_3$ Lie algebra}
\label{sec:sl3}

\paragraph{Roots and weights.} Let us first fix some conventions for the roots and weights of the $\sla_3$ Lie algebra.
The root vectors $\{ \pm\ve_1,\pm\ve_2,\pm(\ve_1+\ve_2) \}$ are the shifts associated to raising and lowering operators. The simple roots are $\{\ve_1,\ve_2\}$. The positive roots $\{\ve_1,\ve_2,\ve_1+\ve_2\}$ are obtained by summing one or sereval distinct simple roots. The dual basis of $(\ve_1,\ve_2)$ is given by the fundamental weights $(\om_1,\om_2)$. We have the relations:
\begin{align}\label{eq:weightsroots}
  & \om_1^2=\om_2^2 = \frac{2}{3} \,, \qquad \om_1\cdot\om_2 = \frac{1}{3} \,, \\
  & \ve_1^2=\ve_2^2 = 2 \,, \qquad \ve_1\cdot\ve_2 = -1 \,, \\
  & \ve_1 = 2\om_1-\om_2 \,, \qquad \ve_2=2\om_2-\om_1 \,, \qquad \ve_i \cdot \om_j = \delta_{ij} \,.
\end{align}
The Weyl vector is $\vrho = \ve_1+\ve_2 = \om_1 + \om_2$.
\bigskip

\paragraph{Irreducible representations.} An irreducible representation (irrep) $[\vlambda]$ is specified by a highest weight vector
\begin{equation}
  \vlambda = (\lambda_1,\lambda_2) = \lambda_1 \om_1 + \lambda_2 \om_2 \,,
  \qquad \lambda_1,\lambda_2 =0,1,2,\dots
\end{equation}
The set of weight vectors of $[\vlambda]$ is constructed recursively, starting from the highest weight $\vlambda$, by the algorithm:
\begin{equation} \label{eq:algo}
  \forall \vlambda'=(\vlambda'_1,\vlambda'_2) \in [\vlambda] \,,
  \quad \text{if } \lambda_i>0 \text{ then }
  \vlambda'-\ve_i, \dots, \vlambda'-\lambda'_i\ve_i \in [\vlambda] \,.
\end{equation}
The multiplicity of the weight $\vlambda'$ in $[\vlambda]$ is denoted $m_{\vlambda}(\vlambda')$, and is obtained by the Freudenthal recursion.

\paragraph{Conjugation.} The conjugate of an irrep is obtained by the reflection around $\vrho$, i.e. the exchange of $\om_1$ and $\om_2$:
\begin{equation}
  (\lambda_1,\lambda_2)^* = (\lambda_2,\lambda_1) \,.
\end{equation}

\paragraph{Some simple representations.} The representations associated to the fundamental weights are three-dimensional. One has
\begin{equation} \label{eq:fundamentalrep}
  [\om_1] = \{ \vh_1, \vh_2, \vh_3 \} \,,
  \qquad [\om_2] = \{ -\vh_1, -\vh_2, -\vh_3 \} \,,
\end{equation}
with
\begin{equation}
  \vh_1 = \om_1 \,, \quad \vh_2=\om_2-\om_1 \,, \quad \vh_3=-\om_2 \,.
\end{equation}
Let us describe two other simple irreps:
\begin{align}
  [\vrho] &= \{\pm \ve_1,\pm \ve_2,\pm\vrho, 0 \} \,, \\
  [2\om_1] &= \{ 2\vh_1,2\vh_2,2\vh_3,-\vh_1,-\vh_2,-\vh_3 \} \,.
\end{align}
The representation $[\vrho]$ has one non-trivial multiplicity: $m_{\vrho}(0)=2$, whereas the weights of $[2\om_1]$ have no degeneracy.

\paragraph{The Weyl group.} The Weyl group $W$ is generated by the reflections about the vectors $\vh_j$. It preserves the set of root vectors. It acts on the $\vh_j$'s as the symmetric group $\mathfrak{S}_3$.
\bigskip

\paragraph{Fusion.} The tensor product of two irreps can be decomposed as a direct sum of irreps:
\begin{equation}
  [\vlambda] \otimes [\vmu] = \bigoplus_{\vect\nu} N_{\vlambda \vmu}^{\vnu} \,.\, [\vnu] \,,
\end{equation}
where the fusion coefficient $N_{\vlambda \vmu}^{\vnu}$ denotes the multiplicity of $[\vect\nu]$ in the decomposition. The $\Zbb_3$ charge of an irrep is defined as the difference:
\begin{equation}
  q_{\vlambda} = \lambda_1-\lambda_2 \,.
\end{equation}
The fusion coefficient obey a $\Zbb_3$ symmetry:
\begin{equation} \label{eq:Z3-sym}
  \text{if}\quad N_{\vlambda \vmu}^{\vnu} \neq 0 \quad \text{then} \quad
  q_{\vlambda} + q_{\vmu} \equiv q_{\vnu} \mod 3 \,.
\end{equation}
Let us give some fusion rules between simple irreps:
\begin{equation} \label{eq:simple-fusion}
  \begin{aligned}
    (1,0) \otimes (1,0) &= (2,0) \oplus (0,1) \,, \\
    (1,0) \otimes (0,1) &= (0,0) \oplus (1,1) \,, \\
    (1,1) \otimes (1,1) &= (0,0) \oplus (2,2) \oplus (0,3) \oplus (3,0) \,, \\
    (2,0) \otimes (2,0) &= (4,0) \oplus (2,1) \oplus (0,2) \,, \\
    (2,0) \otimes (0,2) &= (0,0) \oplus (1,1) \oplus (2,2) \,, \\
    (1,0) \otimes (2,0) &= (3,0) \oplus (1,1) \,.
  \end{aligned}
\end{equation}
Here is a useful identity, valid for any irrep $[\vlambda]$:
\begin{equation} \label{eq:sum-m^2}
  \sum_{\vlambda' \in [\vlambda]} m_{\vlambda}(\vlambda')^2
  = m_{[\vlambda] \otimes [\vlambda^*]}(0) = \sum_{[\wh\vlambda]} N_{\vlambda \vlambda^*}^{\wh\vlambda} m_{\wh\vlambda}(0) \,.
\end{equation}

\noindent\textbf{Proposition:} Let $[\vlambda]$ be an irrep of $\sla_3$. Then:
\begin{itemize}
  \item $[\vlambda]$ includes the weights $\vh_1,\vh_2,\vh_3$ iff $q_{\vlambda}=+1$.
  \item $[\vlambda]$ includes the weights $-\vh_1,-\vh_2,-\vh_3$ iff $q_{\vlambda}=-1$.
  \item $[\vlambda]$ includes the weight $0$ iff $q_{\vlambda}=0$.
\end{itemize}

\noindent\textbf{Proof for $\om_1$:}

\begin{itemize}
\item If $[\vlambda]$ includes the weight $\vh_1=\om_1$, then from the algorithm \eqref{eq:algo}, there exist $(k,\ell) \in \mathbb{N}^2$ such that $\vlambda=\om_1+k\ve_1+\ell\ve_2$, and thus $\lambda_1-\lambda_2 \equiv 1 \mod 3$.

\item If $\lambda_1-\lambda_2 \equiv 1 \mod 3$, then let us prove first that $\om_1 \in [\vlambda]$. From Prop.~1 there exist $(k,\ell) \in \mathbb{N}^2$ such that $\vlambda=\om_1+k\ve_1+\ell\ve_2$. Let us apply the algorithm~\eqref{eq:algo}, starting from the heighest weight $\vlambda$. If $k \geq \ell$ then $\lambda_1\geq k+1>0$, and hence $\vlambda-k\ve_1 \in [\vlambda]$. Now $\vlambda-k\ve_1=(-\ell+1,2\ell)$. If $\ell=0$, then $\vlambda-k\ve_1=\om_1$, and hence $\om_1 \in[\vlambda]$. If $\ell>0$ then $\vlambda-k\ve_1-\ell\ve_2=\om_1 \in[\vlambda]$. A similar argument can be made in the case $k<\ell$.
  Applying \eqref{eq:algo} to $\om_1=\vh_1$, we find that $\vh_1-\ve_1=\vh_2$ and $\vh_2-\ve_2$ also belong to $[\vlambda]$.
  
\end{itemize}

\bibliographystyle{unsrt}
\bibliography{biblio}

\end{document}